\documentclass[journal]{IEEEtran}
\usepackage{amssymb}

\usepackage{subfigure}
\usepackage{graphicx}
\usepackage{enumerate}
\usepackage{amsmath}
\usepackage{amsfonts}
\usepackage{algorithm}
\usepackage{algpseudocode}
\usepackage{url}
\usepackage{epstopdf}
\usepackage{multirow}
\usepackage{authblk}
\usepackage{subfigure}
\usepackage{xcolor}
\usepackage{mathrsfs}

\usepackage{verbatim}

\newtheorem{theorem}{Theorem}[section]
\newtheorem{example}[theorem]{Example}
\newtheorem{corollary}[theorem]{Corollary}
\newtheorem{proposition}[theorem]{Proposition}
\newtheorem{definition}[theorem]{Definition}
\newtheorem{lemma}[theorem]{Lemma}
\newcommand{\prof}{\begin{IEEEproof}}
\newcommand{\eprof}{\end{IEEEproof}}

\newcommand{\prop}{\begin{proposition}}
\newcommand{\eprop}{\end{proposition}}
\newcommand{\them}{\begin{theorem}}
\newcommand{\ethem}{\end{theorem}}
\newcommand{\dfn}{\begin{definition}}
\newcommand{\edfn}{\end{definition}}
\newcommand{\exm}{\begin{example}}
\newcommand{\eexm}{\end{example}}
\newcommand{\coro}{\begin{corollary}}
\newcommand{\ecoro}{\end{corollary}}
\newcommand{\lem}{\begin{lemma}}
\newcommand{\elem}{\end{lemma}}

\newcommand{\eps}{\varepsilon}

\ifCLASSINFOpdf
\else
\fi

\begin{document}

\title{Verification of Detectability Using Petri Nets and Detector}
%
%
%

\author{Hao~Lan, 
        Yin~Tong, ~\IEEEmembership{Member,~IEEE}
        ~Jin~Guo
        and Carla~Seatzu, \IEEEmembership{Senior Member,~IEEE}%
\thanks{H. Lan, Y. Tong (Corresponding Author), and Jin Guo are with the School of Information Science and Technology, Southwest Jiaotong University, Chengdu 611756, China
        {\tt\small haolan@my.swjtu.edu.cn; yintong@swjtu.edu.cn; jguo\_scce@swjtu.edu.cn}}
\thanks{C. Seaztu is with the Department of Electrical and Electronic Engineering,
University of Cagliari, 09123 Cagliari, Italy {\tt\small seatzu@diee.unica.it}}}%

%
%

\markboth{
}%
{Shell \MakeLowercase{\textit{et al.}}: Bare Demo of IEEEtran.cls for IEEE Journals}
%



\maketitle
\begin{abstract}
Detectability describes the property of a system to uniquely determine, after a finite number of observations, the current and subsequent states. In this paper, to reduce the complexity of checking the detectability properties in the framework of bounded labeled Petri nets, we use a new tool, which is called detector, to verifying the strong detectability and periodically strong detectability. First, an approach, which is based on the reachable graph and its detector, is proposed. Then, we develop a novel approach which is based on the analysis of the detector of the basis reachability graph. Without computing the whole reachability space, and without building the observer, the proposed approaches are more efficient.
\end{abstract}


%
\IEEEpeerreviewmaketitle

\section{Introduction}
In recent years, detectability has drawn a lot of attention from researchers in the discrete event system (DES) community \cite{zhang2017problem,yin2017initial,masopust2018complexity, tong2019verification,Lan2019Verification}. This property characterizes the ability of a system to determine the current and the subsequent states of the system after the observation of a finite number of events.
Detectability has been studied earlier in DES with another terminology, which is called ``observability" \cite{ramadge1986observability, ozveren1990observability, giua2002observability}. The observability of the current state and initial state are discussed in \cite{ramadge1986observability}, and whether the current state can be determined periodically is investigated in \cite{ozveren1990observability}.
The property of detectability in DESs has been studied systematically in the literature \cite{zhang2017problem,masopust2018complexity,shu2007detectability,shu2011generalized, keroglou2015detectability}.
The notion of detectability was first proposed and studied in \cite{shu2007detectability} in the deterministic finite automaton framework based on the assumption that the states and the events are partially observable. Shu et al. \cite{shu2007detectability} defined four types of detectability: strong detectability, weak detectability, strong periodic detectability, and weak periodic detectability. And the four types of detectability are verified by an approach whose complexity is exponential with respect to the number of states of the system. Polynomial algorithms to check strong detectability and strong periodic detectability of an automaton have been proposed in \cite{shu2011generalized}. While checking weak detectability and weak periodic detectability is proved to be PSPACE-complete and that PSPACE-hardness \cite{zhang2017problem}, even for a very restricted type of automata \cite{masopust2018complexity}. The notation of detectability is also extended to delayed DESs \cite{shu2013delayed}, modular DESs \cite{yin2017verification} and stochastic DESs \cite{yin2017initial,keroglou2015detectability}, and the enforcement of the detectability is proposed in \cite{shu2013online,yin2016uniform}.

Petri nets are widely used to model many classes of concurrent systems, some problems such as supervisory control \cite{ma2015design}, fault diagnosis \cite{cabasino2011discrete}, opacity \cite{tong2017verification}, etc. can be solved more efficiently in Petri nets.
The detectability of unlabeled Petri nets was proposed by Giua and Seatzu \cite{giua2002observability}, including marking observability and strong marking observability.
In \cite{masopust2019deciding}, the authors extend strong detectability and weak detectability in DESs to labeled Petri nets. Strong detectability is proved to be decidable and checking the property is EXPSPACE-hard, while weak detectability is proved to be undecidable.
In our previous work, we first extend the four detectability properties to labeled Petri nets in \cite{tong2019verification}, then we relax detectability to C-detectability that only requires that a given set of crucial states can be distinguished from other states \cite{Lan2019Verification}.
In \cite{tong2019verification}, it is shown that detectability can be efficiently verified by using Petri nets. However, this method requires the construction of an observer of the \emph{basis reachability graph} (BRG) of the LPN system. Since in the worst case, the complexity of constructing the observer is exponential to the number of states of the BRG. Thus, it is important to search for more efficient algorithms for checking detectability in labeled Petri nets.

In this paper, we develop a method to check strong detectability and periodically strong detectability with lower complexity, compared with \cite{tong2019verification}. By assumption that the initial state of the observed behavior is not known and the systems evolution is only partially observed, the method is based on the construction of a new tool, called ``detector", which was first proposed in \cite{shu2011generalized} for verification of detectability in the framework of automation. We present necessary and sufficient conditions for the strong detectability and periodically strong detectability, by analyzing the detector of the BRG of the original LPN system.
Thanks to basis markings and detector, there is no need to enumerate all the markings and no need to build the observer. This leads to a relevant advantage in terms of computational complexity since the basis reachability graph (BRG) is usually much smaller than the RG and the complexity of the detector is polynomial time.
Further more, rather than computing all cycles in the detector \cite{shu2011generalized,shu2013delayed,shu2012detectability}, which is NP-complete, we show that detectability can be verified with polynomial complexity.


The rest of the paper is organized as follows. In Section~\ref{sec:pre}, background on labeled Petri nets, basis markings and the definition of four detectability properties is provided.
Based on the RG and its detector, we propose an approach to verify the strong detectability, periodically strong detectability in Section~\ref{sec:RG}.
In Section~\ref{sec:BRG}, the efficient approaches to verify the strong detectability, periodically strong detectability are presented.
Conclusions are finally drawn in Section~\ref{sec:con} where our future lines of research in this framework are illustrated.

\section{Preliminaries and Background}\label{sec:pre}
In this section we recall the formalisms used in the paper and some results on state estimation in Petri nets. For more details, we refer to \cite{cabasino2011discrete,murata1989petri,cassandras2009introduction}.

%

\subsection{Petri Nets}\label{subsec:pn}
A \emph{Petri net} is a structure $N=(P,T,Pre,Post)$, where $P$ is a set of $m$ \emph{places}, graphically represented by circles; $T$ is a set of $n$ \emph{transitions}, graphically represented by bars; $Pre:P\times T\rightarrow\mathbb{N}$ and $Post:P\times T\rightarrow\mathbb{N}$ are the \emph{pre-} and \emph{post-incidence functions} that specify the arcs directed from places to transitions, and vice versa. The incidence matrix of a net is denoted by $C=Post-Pre$. A Petri net is \emph{acyclic} if there are no oriented cycles.

A \emph{marking} is a vector $M:P\rightarrow \mathbb{N}$ that assigns to each place a non-negative integer number of tokens, graphically represented by black dots. The marking of place $p$ is denoted by $M(p)$. A marking is also denoted by $M=\sum_{p\in P}M(p)\cdot p$. A \emph{Petri net system} $\langle N,M_0\rangle$ is a net $N$ with \emph{initial marking} $M_0$.

A transition $t$ is \emph{enabled} at marking $M$ if $M\geq Pre(\cdot,t)$ and may fire yielding a new marking $M'=M+C(\cdot,t)$. We write $M[\sigma\rangle$ to denote that the sequence of transitions $\sigma=t_{j1}\cdots t_{jk}$ is enabled at $M$, and $M[\sigma\rangle M'$ to denote that the firing of $\sigma$ yields $M'$. The set of all enabled transition sequences in $N$ from marking $M$ is $L(N,M)=\{\sigma\in T^*| M[\sigma\rangle \}$. Given a transition sequence $\sigma\in T^*$, the function $\pi:T^*\rightarrow \mathbb{N}^n$ associates with $\sigma$ the Parikh vector $y=\pi(\sigma)\in\mathbb{N}^n$, i.e., $y(t)=k$ if transition $t$ appears $k$ times in $\sigma$. Given a sequence of transitions $\sigma\in T^*$, its \emph{prefix}, denoted by $\sigma'\preceq \sigma$, is a string such that $\exists \sigma''\in T^*:\sigma'\sigma''=\sigma$. The \emph{length} of $\sigma$ is denoted by $|\sigma|$.

A marking $M$ is \emph{reachable} in $\langle N,M_0\rangle$ if there exists a transition sequence $\sigma$ such that $M_0[\sigma\rangle M$. The set of all markings reachable from $M_0$ defines the \emph{reachability set} of $\langle N,M_0\rangle$, denoted by $R(N,M_0)$. A Petri net system is \emph{bounded} if there exists a non-negative integer $k \in \mathbb{N}$ such that for any place $p \in P$ and any reachable marking $M \in R(N,M_0)$, $M(p)\leq k$ holds.

A \emph{labeled Petri net} (LPN) system is a 4-tuple $G=(N,M_0,\allowbreak E,\ell)$, where $\langle N,M_0\rangle$ is a Petri net system, $E$ is the \emph{alphabet} (a set of labels) and $\ell:T\rightarrow E\cup\{\eps\}$ is the \emph{labeling function} that assigns to each transition $t\in T$ either a symbol from $E$ or the empty word $\eps$. Therefore, the set of transitions can be partitioned into two disjoint sets $T=T_o\dot{\cup} T_u$, where $T_o=\{t\in T|\ell(t)\in E\}$ is the set of $|T_o|=n_o$ observable transitions and $T_u=T\setminus T_o=\{t\in T|\ell(t)=\eps\}$ is the set of $|T_u|=n_u$ unobservable transitions.
The labeling function can be extended to transition sequences $\ell: T^*\rightarrow E^*$ as $\ell(\sigma t)=\ell(\sigma)\ell(t)$ with $\sigma\in T^*$ and $t\in T$.

Given a set of markings $Y\subseteq R(N,M_0)$, the \emph{language generated by} $G$ \emph{from} $Y$ is ${\cal L}(G,Y)=\bigcup_{M\in Y}\{w\in E^*| \exists \sigma \in L(N,M): w=\ell(\sigma)\}$.
In particular, the \emph{language generated by} $G$ from $R(N,M_0)$ is ${\cal L}(G,R(N,M_0))=\bigcup_{M\in R(N,M_0)}\{w\in E^*| \exists \sigma \in L(N,M): w=\ell(\sigma)\}$ that is simply denoted by ${\cal L}(G)$. Let $w\in {\cal L}(G)$ be an observed word. We denote as
\begin{align}\label{eq:cw2}
{\cal C}(w)=\{M\in \mathbb{N}^m|&\exists M'\in R(N,M_0), \sigma\in L(N,M'): \nonumber \\
& M'[\sigma\rangle M, \ell(\sigma)=w\}.
\end{align}
the set of markings consistent with $w$. When $|{\cal C}(w) | \neq 1$, markings in ${{\cal C}}(w)$ are \emph{confusable} since any of them could be the current marking of the system.
Correspondingly, we denote as $L(G)=\{\sigma\in T^*| \exists M\in R(N,M_0): M[\sigma\rangle\}$ the set of transition sequences enabled at a marking in $R(N,M_0)$. Finally we denote as $L^{\omega}(G)=\{\sigma\in T^*|\sigma\in L(G) \wedge |\sigma|\text{ is infinite}\}$ the set of transition sequences of infinite length that are enabled at some markings in $R(N,M_0)$.


Given an LPN system $G=(N,M_0,\allowbreak E,\ell)$ and the set of unobservable transitions $T_u$, the \emph{$T_u$-induced subnet} $N'=(P,T',\allowbreak Pre',Post')$ of $N$, is the net resulting by removing all transitions in $T\setminus T_u$ from $N$, where $Pre'$ and $Post'$ are the restriction of $Pre$, $Post$ to $T_u$, respectively. The incidence matrix of the $T_u$-induced subnet is denoted by $C_u=Post'-Pre'$.

\subsection{Basis Markings}\label{sec:basis}
In this subsection we review the notion and some results of basis markings, which is proposed in \cite{cabasino2011discrete,tong2017verification,ziyue2017basis}.

\dfn\label{def:exp}
Given a marking $M$ and an observable transition $t\in T_o$, we denote as $$\Sigma(M,t)=\{\sigma\in T^*_u|M[\sigma\rangle M',M'\geq Pre(\cdot,t)\}$$ the set of \emph{explanations} of $t$ at $M$ and $Y(M,t)=\{y_u\in \mathbb{N}^{n_u}|\exists \sigma\in \Sigma(M,t):y_u=\pi(\sigma)\}$ the set of \emph{$e$-vectors}. \hfill $\diamond$
\edfn

After firing any unobservable transition sequence in $\Sigma(M,t)$ at $M$, the transition $t$ is enabled.
To provide a compact representation of the reachability set, we are interested in finding the explanations whose firing vector is minimal.

\dfn\label{def:minexp}
Given a marking $M$ and an observable transition $t\in T_o$, we denote as
$$\Sigma_{min}(M,t)=\{\sigma\in \Sigma(M,t)|\nexists \sigma'\in \Sigma(M,t):\pi(\sigma')\lneqq\pi(\sigma)\}$$
the set of \emph{minimal explanations} of $t$ at $M$ and $Y_{min}(M,t)=\{y_u\in \mathbb{N}^{n_u}|\exists \sigma\in \Sigma_{min}(M,t):y_u=\pi(\sigma)\}$ as the corresponding set of \emph{minimal $e$-vectors}. \hfill $\diamond$
\edfn

There are many approaches to calculate $Y_{min}(M,t)$. In particular, Cabasino \emph{et al} present an approach that only requires algebraic manipulations when the $T_u$-induced subnet is acyclic \cite{cabasino2011discrete}.

\dfn\label{def:basisM}
Given an LPN system $G=(N,M_0,E,\ell)$ whose $T_u$-induced subnet is acyclic, its \emph{basis marking set} ${\cal M}_b$ is defined as follows:
\begin{itemize}
  \item $M_0\in {\cal M}_b$;
  \item If $M\in {\cal M}_b$, then $\forall t\in T_o, y_u\in Y_{min}(M,t)$,
  $$M'=M+C(\cdot,t)+C_u\cdot y_u \Rightarrow M'\in {\cal M}_b.$$
\end{itemize}
A marking $M_b\in {\cal M}_b$ is called a \emph{basis marking} of $G$.
\hfill $\diamond$
\edfn

The set of basis markings contains the initial marking and all other markings that are reachable from a basis marking by firing a transition sequence $\sigma_u t$, where $t\in T_o$ is an observable transition and $\pi(\sigma_u)=y_u$ is a minimal explanation of $t$ at $M$. Clearly, ${\cal M}_b\subseteq R(N,M_0)$, and in practical cases the number of
basis markings is much smaller than the number of reachable markings \cite{cabasino2011discrete,ziyue2017basis}.
And the number of basis markings is finite if the corresponding LPN system is bound.
We denote as ${\cal C}_b(w)={\cal M}_b\cap {\cal C}(w)$ the set of basis markings corresponding to a given observation $w\in {\cal L}(G)$.

\begin{figure}
  \centering
  \includegraphics[width=0.37\textwidth]{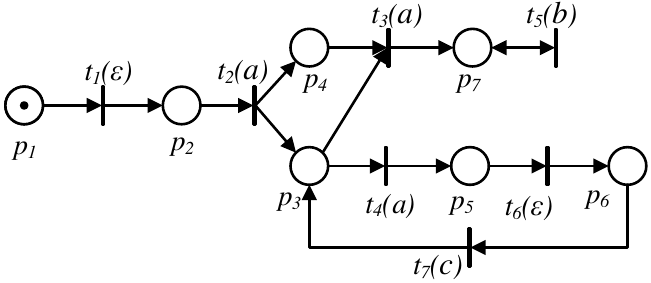}\\
  \caption{The LPN system in Example~\ref{eg:miniExp}.}\label{fig:LPN}
\end{figure}

\exm\label{eg:miniExp}
Let us consider the LPN system in Fig.~\ref{fig:LPN}, where $T_o=\{t_2, t_3, t_4, t_6, t_7\}$, $T_u=\{t_1, t_6\}$. Transitions $t_2$, $t_3$ and $t_4$ are labeled by $a$, transition $t_5$ is labeled by $b$, and transition $t_7$ is labeled by $c$.
At the initial marking $M_0=[1\ 0\ 0\ 0\ 0\ 0\ 0]^T$, the set of minimal explanations of $t_2$ is $\Sigma_{min}(M_0,t_2)=\{t_1\}$, and thus $Y_{min}(M_0,t_2)=\{[1\ 0]^T\}$. The corresponding basis marking is $M_0+C(\cdot,t_2)+C_u\cdot [1\ 0]^T=M_2=[0\ 0\ 1\ 1\ 0\ 0\ 0]^T$.
At $M_2$, the set of minimal explanations of $t_3$ is $\Sigma_{min}(M_2,t_3)=\{\eps\}$, and thus $Y_{min}(M_2,t_3)=\{\vec{0}\}$. The corresponding basis marking obtained is $M_2+C(\cdot,t_3)+C_u\cdot \vec{0}=M_3= [0\ 0\ 0\ 0\ 0\ 0\ 1]^T$.

\hfill $\diamond$
\eexm

\prop\label{prop:basis}\cite{Lan2019Verification,cabasino2011discrete}
Let $G=(N,M_0,E,\ell)$ be an LPN whose $T_u$-induced subnet is acyclic, $M_b\in {\cal M}_b$ a basis marking of $G$, and $w\in {\cal L}(G)$ an observation generated by $G$. We have
\begin{enumerate}
  \item A marking $M$ is reachable from $M_b$ if and only if
\begin{equation}\label{eq:basismarking}
M=M_b+C_u\cdot y_u
\end{equation}
has a nonnegative solution $y_u\in \mathbb{N}^{n_u}$.

\item \begin{align}
{\cal C}(w)=\bigcup_{M_b\in {\cal C}_b(w)}  & \{M\in \mathbb{N}^m|M=M_b+C_u\cdot y_u:  \nonumber\\
  & y_u\in  \mathbb{N}^{n_u}\}  \nonumber
    \end{align}
\end{enumerate}
\eprop

Statement 1) of Proposition~\ref{prop:basis} implies that any solution $y_u \in \mathbb{N}^{n_u}$ of Eq.~\eqref{eq:basismarking} corresponds to the firing vector of a firable transition sequence $\sigma$ from $M_b$, i.e., $M_b[\sigma\rangle$ and $\pi(\sigma)=y_u$. According to Statement 2), the set of markings consistent with an observation can be characterized using
linear algebra without an exhaustive marking enumeration.

\subsection{Detectability}

In this subsection we recall the definitions of the detectability problems of the LPN system.
We assume that the initial marking $M_0$ of the LPN system is given, but the observation could be generated from any marking in $R(N,M_0)$.
As in \cite{tong2019verification}, we make the following two assumptions: 1) the LPN system $G$ is deadlock free. 2) the $T_u$-induced subnet of $G$ is acyclic.
For more details, we refer to \cite{tong2019verification}.

\dfn\label{def:SD}
[\textbf{Strong detectability}]
An LPN system $G=(N,M_0,\allowbreak E,\ell)$ is \emph{strongly detectable}
if
$$\exists K\in \mathbb{N},\forall \sigma\in L^\omega(G), \forall \sigma'\preceq \sigma, |w|\geq K \Rightarrow |{\cal C}(w)|=1,$$
where $w=\ell(\sigma')$. \hfill $\diamond$
\edfn

In words, an LPN system is strongly detectable if the current and the subsequent states of the system can be determined after a finite number of events observed for all trajectories of the system.

\dfn\label{def:WD}
[\textbf{Weak detectability}]
An LPN system $G=(N,M_0,\allowbreak E,\ell)$ is \emph{weakly detectable}
if
$$ \exists K\in \mathbb{N},\exists \sigma\in  L^\omega(G), \forall \sigma'\preceq \sigma, |w|\geq K \Rightarrow |{\cal C}(w)|=1,$$
where $w=\ell(\sigma')$. \hfill $\diamond$
\edfn

In simple words, an LPN system is weakly detectable if we can determine, after a finite number of observations,
the current and subsequent states of the system for some trajectories of the system.

\dfn\label{def:SPD}
[\textbf{Periodically strong detectability}]
An LPN system $G=(N,M_0,\allowbreak E,\ell)$ is \emph{periodically strongly detectable} if
$\exists K\in \mathbb{N}, \forall \sigma\in L^\omega (G),\forall \sigma' \preceq \sigma$,
$$  \exists \sigma''\in T^*: \sigma'\sigma''\preceq \sigma \wedge |\ell(\sigma'')|\leq K \Rightarrow |{\cal C}(w)|=1,$$
where $w=\ell(\sigma'\sigma'')$. \hfill $\diamond$
\edfn

Therefore, an LPN system is periodically strongly detectable if the current and the subsequent states of the system can be periodically determined for all trajectories of the system.

\dfn\label{def:WPD}
[\textbf{Periodically weak detectability}]
An LPN system $G=(N,M_0,\allowbreak E,\ell)$ is \emph{periodically weakly detectable}
if
$\exists K\in \mathbb{N}$, $\exists \sigma\in L^\omega (G),\forall \sigma' \preceq \sigma$,
$$  \exists \sigma''\in T^*: \sigma'\sigma''\preceq \sigma \wedge |\ell(\sigma'')|\leq K \Rightarrow |{\cal C}(w)|=1,$$
where $w=\ell(\sigma'\sigma'')$. \hfill $\diamond$
\edfn

In words, an LPN system is periodically weakly detectable if we can periodically determine the current state of the system for some trajectories of the system.

\begin{figure}
  \centering
  \subfigure[]{%
  \label{fig:RG}
  \includegraphics[width=0.5\textwidth]{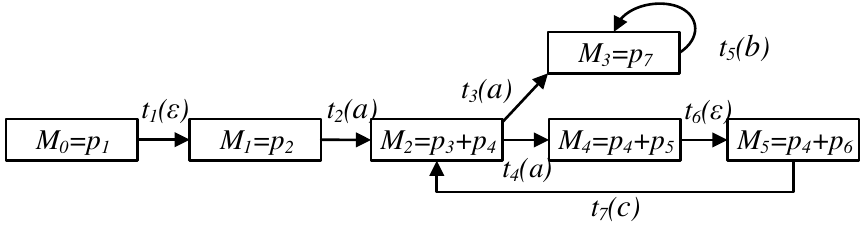}}
  \centering
  \subfigure[]{%
  \label{fig:obs}
  \includegraphics[width=0.37\textwidth]{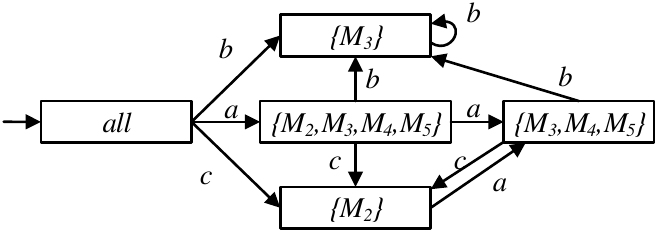}}
  \caption{The RG of the LPN system in Fig.~\ref{fig:LPN} (a), the observer of the RG (b).}
  \label{fig:WD}
\end{figure}

\exm\label{eg:Obsr}
Let us consider again the LPN system in Fig.~\ref{fig:LPN}. Its RG is shown in Fig.~\ref{fig:RG}, and the observer of RG is shown in Fig.~\ref{fig:obs}.
When $(ac)^*$ is observed (the LPN system fires $(t_4t_6t_7)^*$), the current state of the system can be uniquely determined, that is $M_2$.
However, there always exists two arbitrarily long prefix $(t_4t_6t_7)^*t_4$ and $(t_4t_6t_7)^*t_3$ (they have the same observation $(ac)^*a$) such that the current state cannot be determined, that is, if $(ac)^*a$ is observed, the current state could be any state in $\{ M_3,M_4,M_5\}$. Therefore, according to Definitions~\ref{def:SD}, the LPN system is not strongly detectable.

On the other hand, when the LPN system fires $(t_4t_6t_7)^*$ and we observe $(ac)^*$, we know that the current state of the system is $M_2$ periodically (after seeing $c$). And when observing $(b)^*$ (the LPN system fires $(t_5)^*$), $M_3$ is the current state of the system.
Therefore, according to Definitions~\ref{def:WD}, \ref{def:SPD} and \ref{def:WPD}, the LPN system is weakly detectable, periodically strongly detectable and periodically weakly detectable. \hfill $\diamond$
\eexm


\section{RG and its detector}\label{sec:RG}

In automation framework, detector was proposed to verifying the strong detectability and periodically strong detectability \cite{shu2011generalized,shu2013delayed,shu2012detectability}. As \cite{shu2011generalized} shows that the complexity of construction of the detector is polynomial with respect to the number of states of the system, which is lower than the observer. Obviously, the same approach can be used on the bounded LPN system.
Thus, in this section, we construct the detector of the RG of the bounded LPN system, to check the strong detectability and periodically strong detectability of the LPN system.

As in \cite{shu2011generalized}, the detector of the RG is denoted by
$$D=(Q, E, f_{r}, q_{0}),$$ where $Q \subseteq 2^{R(N,M_0)}$ is a finite set of states.
Since it is assumed that the marking from which the observation is generated is not known, the initial state of $D$ is $q_{0}=R(N,M_0)$, and the other states of $D$ is $q\subseteq R(N,M_0) \wedge |q|\leq 2$. The event set of the detector is the alphabet $E$.
We denote as $UR(M)=\{M'\in \mathbb{N}^{m} | M[\sigma_u\rangle M',\sigma_u\in T^*_u\}$
the unobservable reach of the marking $M$. 
The transition function $f_{r}:Q\times E \rightarrow 2^{Q}$ is defined as follows.

Given a state $q \subseteq R(N,M_0), e\in E$, $t\in T, \ell(t)=e$. Let
$q_t=UR(\{M\in R(N,M_0)| \exists M'\in q, M'[t\rangle M\}),$ then,
$$
f_r(q,e)=\left\{\begin{array}{ll}
  \{q_t\} & \text{if $|q_t|=1$;}\\
  \{q'|q'\subseteq q_t \wedge |q'|=2\} & \text{if $|q_t|\geq 2$;}\\
  undefined & \text{otherwise.}
\end{array}\right.
$$

\begin{figure}
  \centering
  \includegraphics[width=0.35\textwidth]{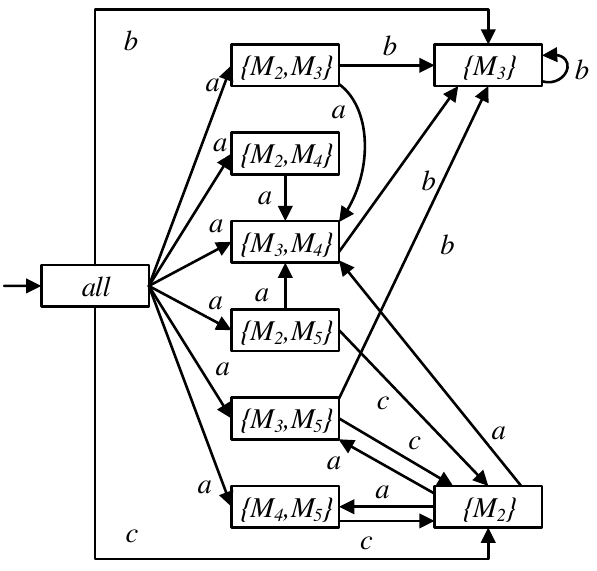}\\
  \caption{The detector of the RG in Fig.~\ref{fig:RG}.}\label{fig:deRG}
\end{figure}

\exm\label{eg:deRG}
Consider again the LPN system in Fig.~\ref{fig:LPN}, its RG is shown in Fig.~\ref{fig:RG}. By the construction method, the detector of the BRG is presented in Fig.~\ref{fig:deRG}. The initial state is all the markings of the RG in Fig.~\ref{fig:RG}. When $a$ is observed at the initial state, there are four markings may be reached in the RG. Thus according to the construction method, the initial state can reach six states with a combination of the four markings.
\hfill $\diamond$
\eexm

Essentially, the detector of RG is constructed by splitting and recombining the state in ${\cal C}(w)$ when ${\cal C}(w)$ contains more than two elements. Namely, for any state $q=f_r(M_0,w)$ in $D$, $q \subseteq {\cal C}(w)$.

Since detectability considers the transition sequences of infinite length, we first study the properties of cycles in the detector of the RG.

\dfn\label{def:RGcycle}
A \emph{(simple) cycle} in the detector $D=(Q, E, f_{r}, q_{0})$ of a RG is a path $\gamma_j=q_{j1}e_{j1}q_{j2}\ldots q_{jk}\allowbreak e_{jk}q_{j1}$ that starts and ends at the same state but without repeated edges, where $q_{ji}\in Q$ and $e_{ji}\in E$ with $i=\{1,2,\ldots,k\}$, and $\forall m,n\in\{1,2,\ldots,k\}$ with $m\neq n$, $q_{jm}\neq q_{jn}$. The corresponding observation of the cycle is $w=e_{j1}\ldots e_{jk}$. A state $q_{ji}$ contained in $\gamma_j$ is denoted by $q_{ji}\in \gamma_j$. 
\hfill $\diamond$
\edfn

Since RG is actually an automation, thus, we can conclude the following theorem according to \cite{shu2011generalized}.

\them\label{therm:SDRG}
Let $G=(N,M_0,\allowbreak E,\ell)$ be an LPN system whose $T_u$-induced subnet is acyclic, and $D=(Q, E, f_{r}, q_{0})$ the detector of its RG. The LPN system $G$ is strongly detectable iff for any $q \in Q$ reachable from a cycle in $D$, it is $|q|=1$. 
\ethem

In words, an LPN system is strongly detectable if and only if in the detector of the RG, such that all the states reachable from any cycle that the cardinality of these states is 1.

\them\label{therm:PSDRG}
Let $G=(N,M_0,\allowbreak E,\ell)$ be an LPN system whose $T_u$-induced subnet is acyclic, and $D=(Q, E, f_{r}, q_{0})$ the detector of its RG. The LPN system $G$ is periodically strongly detectable iff for any cycle $\gamma_j$ in $D$, $\exists q \in \gamma_j$, $|q|=1$. 
\ethem
%

In words, an LPN system is periodically strongly detectable if and only if in the detector of the BRG, such that all the cycles have a state whose cardinality is 1.

\textbf{Remark 1}:
Although the construction of the detector according to \cite{shu2011generalized} is polynomial time complexity, it is known that the complexity of finding all the cycles in a directed graph is NP-complete. Thus, the complexity of the detector based approaches proposed in \cite{shu2011generalized} is not actually polynomial time.
However, finding all the strongly connected components (SCC) is of polynomial complexity w.r.t the size of the graph. Clearly, if a state of the observer is reachable from a cycle, it is also reachable from an SCC. Therefore, Theorem~\ref{therm:SDRG} can be rephrased as follows.

\coro\label{coro:SDRG}
Let $G=(N,M_0,\allowbreak E,\ell)$ be an LPN system whose $T_u$-induced subnet is acyclic, and $D=(Q, E, f_{r}, q_{0})$ the detector of its RG. The LPN system $G$ is strongly detectable iff for any $q \in Q$ reachable from an SCC in $D$, it is $|q|=1$. 
\ecoro

\textbf{Remark 2}:
According to Theorem~\ref{therm:PSDRG}, we also need to check all the cycles and we cannot
take advantage from the usage of SCCs. However, it is easy to find that we can check Theorem~\ref{therm:PSDRG} by its contrapositive. Thus we just need to find one cycle according to the following corollary, which makes the approach polynomial complexity.

\coro\label{coro:PSDRG}
Let $G=(N,M_0,\allowbreak E,\ell)$ be an LPN system whose $T_u$-induced subnet is acyclic, and $D=(Q, E, f_{r}, q_{0})$ the detector of its RG. The LPN system $G$ is not periodically strongly detectable iff there exists a cycle $\gamma_j$ in $D$, for any states $q \in \gamma_j$, $|q|\neq 1$.
\ecoro


\exm\label{eg:ver-deRG}
Consider again the LPN system in Fig.~\ref{fig:LPN}. Its RG is shown in Fig.~\ref{fig:RG}, and the detector of the RG is shown in Fig.~\ref{fig:deRG}. Now we use Theorem~\ref{therm:SDRG} and \ref{therm:PSDRG} to check its strong detectability and periodically strong detectability.
In the detector of RG, we can see that there is a cycle $\gamma_1=\{M_2\}a\{M_4,M_5\}c\{M_2\}$ containing state $\{M_4,M_5\}$ whose cardinality is 2, thus, there exists a cycle that does not satisfy all states $|q|=1$. Therefore, the LPN system is not strongly detectable.

On the other hand, the state $\{M_2\}$ in $\gamma_1$ satisfy $|q|=1$.
And we cannot find a cycle that all its states is $|q|\neq 1$. Therefore, according to Corollary~\ref{coro:PSDRG}, the LPN system is periodically strongly detectable.
\hfill $\diamond$
\eexm

\section{BRG and its detector}\label{sec:BRG}

Checking detectability properties based on Theorem~\ref{therm:SDRG} to \ref{therm:PSDRG} (or Corollary~\ref{coro:SDRG} to \ref{coro:PSDRG}) requires the construction of a RG and its detector. It is known that, the complexity of constructing the RG of a Petri net system is exponential in the size of the net (number of places, transitions, tokens in the initial marking).
Therefore, to verify the detectability of large dimension systems, such an approach may not be feasible.

In our previous work \cite{tong2019verification,Lan2019Verification}, we show how the above four detectability properties can be verified using the notion of basis marking and observer, thus avoiding an exhaustive enumeration of all the states in the RG. In this way, the state explosion problem is practically avoided \cite{tong2015verification}. However, the step of building the observer is exponential complexity in the worst case.

Since the BRG is usually much smaller than the RG and the complexity of constructing the detector is lower than the observer, thus, in this subsection, we build the BRG of the LPN system and explore the detector of the BRG to verifying strong detectability and periodically strong detectability.
\subsection{BRG}
Using the notion of basis marking, we introduce the basis reachability graph (BRG) for detectability. To guarantee that the BRG is finite, we assume that the LPN system is bounded. For each basis marking $M_b \in {\cal M}_b$ a binary scalar is assigned by function $\Psi(M_b) : {\cal M}_b \rightarrow \{0,1\}$
that are defined by Eqs.~\eqref{eq:basis}:

\begin{equation}\label{eq:basis}
\Psi(M_b)=\left\{\begin{array}{ll}
  1 & \text{if $M_b+C_u\cdot y_u\geq 0$ has a}\\
    & \text{positive integer solution;}\\
  0 & \text{otherwise.}
\end{array}\right.
\end{equation}

We denote as $B=(X,E,f,x_0)$ the BRG for detectability of an LPN system $G=(N,M_0,E,\ell)$, where $X\in {\cal M}_b \times \{0,1\}$ is a finite set of states, and each state $x\in X$ of the BRG is a pair $(M_b,\Psi(M_b))$. We denote as $x(1)$, $x(2)$ the first and the second element of $x$ respectively. The initial state of the BRG is $x_0=(M_0,\Psi(M_0))$. The event set of the BRG is identical to the alphabet $E$. The transition relation $f: X \times E \rightarrow X$ can be determined by the following rule. If at marking $M_{b}$ there is an observable transition $t$ for which a minimal explanation exists and the firing of $t$ and one of its minimal explanations leads to $M_{b}'$, then an edge from node $(M_{b},\Psi(M_b))$ to node $(M_b',\Psi(M_b'))$ labeled with $\ell(t)$ is defined in the BRG. The procedure to construct the BRG for detectability is summarized in Algorithm~1 in \cite{tong2019verification}.

\begin{figure}
  \centering
  \includegraphics[width=0.35\textwidth]{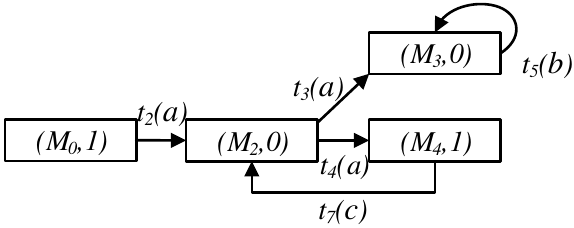}\\
  \caption{The BRG of the LPN system in Fig.~\ref{fig:LPN}.}\label{fig:BRG}
\end{figure}

\exm\label{eg:BRG}
Let us consider again the LPN system in Fig.~\ref{fig:LPN} whose $T_u$-induced subnet is acyclic.
The LPN system has 6 reachable markings and only 4 of them are basis markings, namely, $M_0,M_2,M_3, M_4$. When $M_b$ in Eq.~\eqref{eq:basis} equals $M_0$, the equation has one positive integer solution. Thus, $\Psi(M_0)=1$. On the other hand, for $M_2$, Eq.~\eqref{eq:basis} does not have a positive solution. Therefore, $\Psi(M_2)=0$. The same for other basis markings, thus, according to Algorithm~1 in \cite{tong2019verification}, the BRG for detectability is the graph in Fig.~\ref{fig:BRG}. Note that in Fig.~\ref{fig:BRG} no initial state is pointed out since the initial state is assumed to be unknown.  \hfill $\diamond$
\eexm

\lem\label{lem:psi}\cite{tong2019verification}
Let $G$ be an LPN system whose $T_u$-induced subnet is acyclic, $M_b\in {\cal M}_b$ a basis marking of $G$. If $\Psi(M_b)=1$, there exists an observation $w$ such that $|{\cal C}(w)|\neq 1$.
\elem

In a simple word, if $\Psi(M_b)=1$, then there exists an observation $w$ such that $|{\cal C}(w)|$ contains more than one marking. However, even if $\Psi(M_b)=0$ there may be another basis marking $M_b'$ such that $M_b,M_b'\in {\cal C}(w)$. In this case, $|{\cal C}(w)|$ is still not equal to 1.

In the following, we construct the detector of the BRG to derive necessary and sufficient conditions for detectability.

\subsection{detector of the BRG}

We construct a detector of the BRG $B=(X,E,f,x_0)$ for detectability as in \cite{shu2011generalized}:

$$B_d=(Q_d, E, f_{d}, q_{d0}),$$ where $Q_d \subseteq 2^X$ is a finite set of states. The initial state of $B_d$ is $q_{d0}=X$, and the other states of $B_d$ is $q_d\subseteq X \cap |q_d|\leq 2$. The event set of the detector is the alphabet $E$. The transition function $f_{d}:Q_d\times E \rightarrow 2^{Q_d}$ is defined as follows.

Given a state $q_d \subseteq X, e\in E$, let $q_t=\{x\in X| \exists x'\in q_d, x\in f(x',e)\}$, then,
$$
f_d(q_d,e)=\left\{\begin{array}{ll}
  \{q_t\} & \text{if $|q_t|=1$;}\\
  \{q_d'|q_d'\subseteq q_t \wedge |q_d'|=2\} & \text{if $|q_t|\geq 2$;}\\
  undefined & \text{otherwise.}
\end{array}\right.
$$

\begin{figure}
  \centering
  \includegraphics[width=0.45\textwidth]{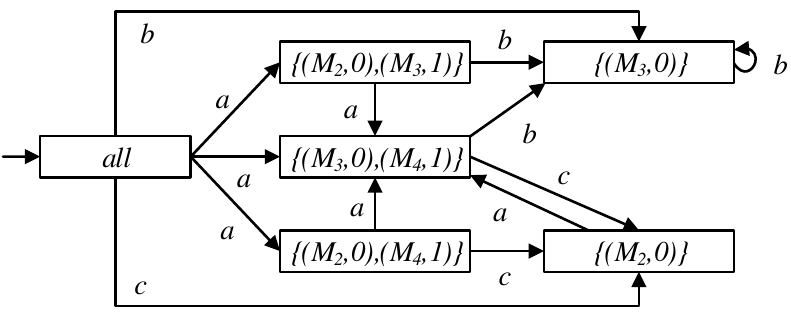}\\
  \caption{The detector of the BRG in Fig.~\ref{fig:BRG}.}\label{fig:deBRG}
\end{figure}

\exm\label{eg:deBRG}
Consider again the LPN system in Fig.~\ref{fig:LPN}, its BRG is shown in Fig.~\ref{fig:BRG}. By the construction method, the detector of the BRG is presented in Fig.~\ref{fig:deBRG}. The initial state is all the basis markings of the BRG in Fig.~\ref{fig:BRG}. When $a$ is observed at the initial state, there are three basis markings may be reached in the BRG. Thus according to the construction method, the initial state can reach three states with a combination of the three basis markings.
\hfill $\diamond$
\eexm

Essentially, the detector of BRG is constructed by splitting and recombining the state in ${\cal C}_b(w)$. When $|{\cal C}_b(w)|>2$, the detector pairs all states in ${\cal C}_b(w)$ in groups of tow.
Namely, for any state $q_d= f_d(M_0,w)$ in $B_d$, $\bigcup_{x\in q_d} x(1) \subseteq {\cal C}_b(w) \subseteq {\cal C}(w)$.

%
%

\lem\label{lem:BRG1}
Let $G$ be an LPN system whose $T_u$-induced subnet is acyclic, and $B_d=(Q_d, E, f_{d}, q_{d0})$ the detector of its BRG.
If there exists a state $q_d\in Q_d$ such that $|q_d|=2$, then there exists an observation $w\in E$ such that $|{\cal C}(w)| \neq 1$.
\elem
\prof
Since by assumption $|q_d|=2$, let $q_d=\{x_1,x_2\}$, $x_1 \neq x_2$. According to the construction of the detector of BRG, then there must exists an observation $w$ such that $f_d(M_0,w)= q_d = \{x_1, x_2\}$, $x_1 \neq x_2$. Thus $x_1(1), x_2(2) \in {\cal C}(w)$. Therefore, $|{\cal C}(w)| \neq 1$.
\eprof

In a simple word, if $|q_d|=2$, then there exists an observation $w$ such that ${\cal C}(w)$ contains more than one marking.

\lem\label{lem:BRG2}
Let $G$ be an LPN system whose $T_u$-induced subnet is acyclic, and $B_d=(Q_d, E, f_{d}, q_{d0})$ the detector of its BRG.
if there exists a state $q_d\in Q_d$ such that $\exists x\in q_d$ that $x(2)=1$, then there exists an observation $w\in E$ such that $|{\cal C}(w)| \neq 1$.
\elem
\prof
Follow from Lemma~\ref{lem:psi}.
\eprof

In a simple word, if $\exists x\in q_d$ that $x(2)=1$, then there exists an observation $w$ such that ${\cal C}(w)$ contains more than one marking.

\prop\label{prop:BRG3}
Let $G$ be an LPN system whose $T_u$-induced subnet is acyclic, and $B_d=(Q_d, E, f_{d}, q_{d0})$ the detector of its BRG.
There exists an observation $w\in E$ such that $|{\cal C}(w)| \neq 1$, iff there exists a state $q_d\in Q_d$ such that $|q_d|=2$ or $\exists x\in q_d$ that $x(2)=1$.
\eprop
\prof
(If) Follow from Lemma~\ref{lem:BRG1} and Lemma~\ref{lem:BRG2}.

(Only if) Assume that there exists an observation $w\in E$ such that $|{\cal C}(w)| \neq 1$, thus there exists two different markings $M_1,M_2 \in {\cal C}(w)$ with $M_1 \neq M_2$. According to the construction of the detector of BRG,
if $M_1,M_2 \in {\cal C}_b(w)$, thus there must exist a state $q_d\in Q_d$ such that $|q_d|=2$;
if $M_1,M_2$ not all in ${\cal C}_b(w)$, thus there must exist a state $q_d\in Q_d$ such that $\exists x\in q_d$ that $x(2)=1$.
\eprof

In words, in an LPN system, there exists an observation $w$ such that ${\cal C}(w)$ contains more than one marking, if and only if there exists a state $q_d$ such that $|q_d|=2$ or $\exists x\in q_d$ that $x(2)=1$.

\coro
Let $G$ be an LPN system whose $T_u$-induced subnet is acyclic, and $B_d=(Q_d, E, f_{d}, q_{d0})$ the detector of its BRG.
If $\forall q_d\in Q_d$, $|q_d|=2$ or $\exists x\in q_d$ that $x(2)=1$.
the LPN system $G$ does not satisfy any detectability property.
\ecoro

\prop\label{prop:BRG4}
Let $G$ be an LPN system whose $T_u$-induced subnet is acyclic, and $B_d=(Q_d, E, f_{d}, q_{d0})$ the detector of its BRG.
If there exists an observation $w\in E$ such that $|{\cal C}(w)| = 1$, then there exists a state $q_d\in Q_d$ such that $q_d=\{(M_b,0)\}$, where $M_b\in {\cal M}_b$.
\eprop
\prof
Since by assumption $|{\cal C}(w)|=1$, let ${\cal C}(w)=\{M_b\}$, according to the construction of the detector of BRG, then $\bigcup_{x\in q_d} x(1)={\cal C}(w)=\{M_b\}$. Thus, $|\bigcup_{x\in q_d} x(1)|=1$, i.e, there is only one state in $q_d$. Since $|{\cal C}(w)|=1$, by Lemma~\ref{lem:psi}, $x(2)=0$. Therefore, $q_d=\{(M_b,0)\}$.
\eprof

In words, if there exists an observation $w$ such that ${\cal C}(w)$ contains only one marking, then the corresponding state $q_d$ in $B_d$ contains only one basis marking $M_b$ and $\Psi(M_b)=0$.
However, the converse is not true.

Similar to Section~\ref{sec:RG}, we denote the simple cycles in the detector of the BRG as follows:

A \emph{(simple) cycle} in the detector $B_d=(Q_d, E, f_{d}, q_{d0})$ of a BRG is a path $\tau_j=q_{j1}e_{j1}q_{j2}\ldots q_{jk}\allowbreak e_{jk}q_{j1}$ that starts and ends at the same state but without repeated edges, where $q_{ji}\in Q_d$ and $e_{ji}\in E$. The corresponding observation of the cycle is $w=e_{j1}\ldots e_{jk}$.

%

\them\label{therm:SD}
Let $G$ be an LPN system whose $T_u$-induced subnet is acyclic, and $B_d=(Q_d, E, f_{d}, q_{d0})$ the detector of its BRG. The LPN system $G$ is strongly detectable iff for any $q_d \in Q_d$ reachable from a cycle in $B_d$, it is $q_d=\{(M_b,0)\}$, where $M_b\in {\cal M}_b$.
\ethem
\prof
Please see Appendix A for the proof.
\eprof

In words, an LPN system is strongly detectable if and only if in the detector of the BRG, such that all the states reachable from any cycle have the form $\{(M_b,0)\}$, i.e., there is only one element $(M_b,\Psi(M_b))$ in these states and $\Psi(M_b)=0$.


According to Remark 1, we can also take advantage from the usage of SCCs. Thus, Theorem~\ref{therm:SD} can be rephrased as follows.

\coro
Let $G$ be an LPN system whose $T_u$-induced subnet is acyclic, and $B_d=(Q_d, E, f_{d}, q_{d0})$ the detector of its BRG. The LPN system $G$ is strongly detectable iff for any $q_d \in Q_d$ reachable from an SCC in $B_d$, it is $q_d=\{(M_b,0)\}$, where $M_b\in {\cal M}_b$.
\ecoro

\them\label{therm:PSD}
Let $G$ be an LPN system whose $T_u$-induced subnet is acyclic, and $B_d=(Q_d, E, f_{d}, q_{d0})$ the detector of its BRG. The LPN system $G$ is strongly periodically detectable iff for any cycle $\tau_j$ in $B_d$, $\exists q_d \in \tau_j$, $q_d=\{(M_b,0)\}$, where $M_b\in {\cal M}_b$.
\ethem
\prof
Please see Appendix B for the proof.
\eprof

In words, an LPN system is periodically strongly detectable if and only if in the detector of the BRG, such that all the cycles have a state having the form $\{(M_b,0)\}$.

By Remark 2, Theorem~\ref{therm:PSD} can also be written as follows.

\coro
Let $G$ be an LPN system whose $T_u$-induced subnet is acyclic, and $B_d=(Q_d, E, f_{d}, q_{d0})$ the detector of its BRG. The LPN system $G$ is not strongly periodically detectable iff there exists one cycle $\tau_j$ in $B_d$, for all $q_d \in \tau_j$, $q_d \neq \{(M_b,0)\}$, where $M_b\in {\cal M}_b$.
\ecoro


\exm\label{eg:ver-deBRG}
Consider again the LPN system in Fig.~\ref{fig:LPN}. Its BRG is shown in Fig.~\ref{fig:BRG}, and the detector of the BRG is shown in Fig.~\ref{fig:deBRG}. Now we use Theorem~\ref{therm:SD} and \ref{therm:PSD} to check its strong detectability and periodically strong detectability.
In Fig.~\ref{fig:deBRG}, we can see that there is a cycle $\tau_1=\{(M_2,0)\}a\{(M_3,0),(M_4,1)\}c\{(M_2,0)\}$ containing state $\{(M_3,0),(M_4,1)\}$ whose cardinality is 2 and $\Psi(M_4)=1$, thus, there exists a cycle that does not satisfy all states $q_d=\{(M_b,\Psi(M_b))\}$ with $\Psi(M_b)=0$. Therefore, the LPN system is not strongly detectable.

On the other hand, the state $\{(M_2,0)\}$ in $\tau_1$ satisfy the form $\{(M_b,0)\}$. And in another cycle $\tau_2=\{(M_3,0)\}c\{(M_3,0)\}$, the only state $\{(M_3,0)\}$ also satisfy the form $\{(M_b,0)\}$, therefore, the LPN system is periodically strongly detectable.

\hfill $\diamond$
\eexm

\section{Conclusion and future work}\label{sec:con}
In this paper, a novel approach to verifying detectability of bounded labeled Petri nets is developed. Our approach is based on the basis marking, and on the exploration of its detector for the detectability. For Petri nets whose unobservable subnet is acyclic, the strong detectability and periodically strong detectability property can be decided by just constructing the detector of the BRG. Since a complete enumeration of possible firing transition sequences is avoided and there is no need for the construction of observer, the proposed approach is of lower complexity than the previous approaches.
The future research is to study on an algorithm that can check the weak detectability and periodically weak detectability with low complexity.

\section*{Acknowledgment}
This work was supported by the National Natural Science Foundation of China under Grant No. 61803317,
the Fundamental Research Funds for the Central Universities under Grant No. 2682018CX24, the Sichuan Provincial S\&T Innovation Project under Grant No. 2018027, and the Key program for International S\&T Cooperation of Sichuan Province under Grant No. 2019YFH0097.

\ifCLASSOPTIONcaptionsoff
  \newpage
\fi



%
%
%

\bibliographystyle{IEEEtran}
\bibliography{detector}

\section*{Appendices}

\subsection{Proof of Theorem~\ref{therm:SD}}\label{app:SD}
(If)
Assume LPN system $G$ is not strongly detectable, that is for all $K\in \mathbb{N}$, there exist $\sigma\in L^\omega(G)$ and $\sigma'\preceq \sigma, |w|\geq K \Rightarrow |{\cal C}(w)|\neq 1$, where $w'=\ell(\sigma')$.
Since $\sigma$ is of an infinite length and $B_d$ has a finite number of nodes, the path along $\ell(\sigma)=w$ must contain a cycle $\tau_j=q_{j1}e_{j1}q_{j2}\ldots q_{jk}\allowbreak e_{jk}q_{j1}$, i.e., there exist $w_0,w_2\in E^*$, such that $w=w_0(e_{j1}\ldots e_{jk})^*w_2$ where $|w_0|$ is finite.
Since the $T_u$-induced subnet is acyclic, let a prefix $\sigma'$ of $\sigma$, $\ell(\sigma')=w_0w''$, $|\ell(\sigma')|\geq K$, where $w'' \preceq (e_{j1}\ldots e_{jk})^*w_2$. Let $q_d=f_d(q_{d0},w_0w'')$, since $|{\cal C}(w)|\neq 1$, $w'=\ell(\sigma')=w_0w''$, by proposition~\ref{prop:BRG3}, thus $|q_d|=2$ or $\exists x\in q_d$ that $x(2)=1$. Namely, there exists a state $q_d$ reachable from a cycle in $B_d$, it is $|q_d|=2$ or $\exists x\in q_d$ that $x(2)=1$.

(Only if)
Assume there exists a cycle $\tau_j$ in $B_d$, $q_r \in \tau_j, w'\in E^*$, such that $q_d=f_d(q_r,w')$ is defined and $|q_d|=2$ or $\exists x\in q_d$ that $x(2)=1$.
Clearly, $B_d$ has a finite number of nodes, then there exists an observation $w_0$ such that $q_r =f_d(q_{d0},w_0)$, with $|w_0|$ is finite. Since $q_r \in \tau_j$, then there exists $\sigma\in L^\omega(G)$ and $w=\ell(\sigma)$, $w_0,w_2\in E^*$ such that $w=w_0(e_{j1}\ldots e_{jk})^*w_2$.
Since the $T_u$-induced subnet is acyclic, then there exist $\sigma'\preceq\sigma$ with $\ell(\sigma')=w_0w''$, $|\ell(\sigma')|\geq K$, where $w'' \preceq (e_{j1}\ldots e_{jk})^*w_2$ and $f_d(q_{d0},w_0)=q_r$. By assumption $q_d=f_d(q_r,w')$ is defined and $|q_d|=2$ or $\exists x\in q_d$ that $x(2)=1$, and $f_d(q_{d0},w_0w')=f_d(q_r,w')=q_d$, thus by Proposition~\ref{prop:BRG3}, this implies that the implication $|{\cal C}(w_0w')|\neq1$ holds.

\subsection{Proof of Theorem~\ref{therm:PSD}}\label{app:PSD}
(If)
Assume LPN system $G$ is not periodically strongly detectable, that is for all $K\in \mathbb{N}$, there exist $\sigma\in L^\omega(G)$ and $\sigma'\preceq \sigma$, $\forall \sigma''\in T^*, \ell(\sigma'\sigma'')=w': \sigma'\sigma''\preceq \sigma, |\ell(\sigma'')|\leq K \Rightarrow |{\cal C}(w')| \neq 1$.
Since $\sigma$ is of an infinite length and $B_d$ has a finite number of nodes, the path along $\ell(\sigma)=w$ must contain a cycle $\tau_j=q_{j1}e_{j1}q_{j2}\ldots q_{jk}\allowbreak e_{jk}q_{j1}$, i.e., there exist $w_0\in E^*$ such that $w=w_0(e_{j1}\ldots e_{jk})^*$ where $|w_0|$ is finite. Let $\ell(\sigma')=w_0, \ell(\sigma'')=w'' \preceq (e_{j1}\ldots e_{jk})^*$. Since $|\ell(\sigma'')|\leq K$, any state $f_d(q_{d0},w')= f_d(q_{d0},w_0w'')=q_{jr}$ must in the cycle $\tau_j$.
Since $|{\cal C}(w')|\neq 1$, by Proposition~\ref{prop:BRG3}, it is $|q_{jr}|=2$ or $\exists x\in q_{jr}$ that $x(2)=1$.

(Only if)
Assume there exists cycle $\tau_j=q_{j1}e_{j1}q_{j2}\ldots q_{jk}\allowbreak e_{jk}q_{j1}$ in $B_d$, $\forall q_{jr} \in \tau_j$, $|q_{jr}|=2$ or $\exists x\in q_{jr}$ that $x(2)=1$.
Clearly, there exist $\sigma\in L^\omega(G)$ and $w_0 \in E^*$, such that $\ell(\sigma)=w=w_0 (e_{j1}e_{j2}\ldots e_{jk})^*$ with $|w_0|$ is finite. Since the $T_u$-induced subnet is acyclic, then there exists $\sigma_0$ with $\ell(\sigma_0)=w_0$, for all $\sigma_1 \in T^*$ with $\ell(\sigma_1)=w_1 \preceq (e_{j1}e_{j2}\ldots e_{jk})^*$, $f_d(q_{d0}, w_0w_1)=q_{jr} \in \tau_j$. By assumption $|q_{jr}|=2$ or $\exists x\in q_{jr}$ that $x(2)=1$, therefore, by Proposition~\ref{prop:BRG3}, $|{\cal C}(w_0w_1)|\neq1$.

\end{document}